\documentclass[12pt,floatfix]{JHEP3}
\usepackage{graphicx}
\usepackage{dcolumn}
\usepackage{bm}
\usepackage{amsfonts}
\usepackage{amsthm}
\usepackage{amsmath}
\usepackage{amssymb}
\usepackage{cite}
\usepackage{epsfig}

\newcommand{\bea}{\begin{eqnarray}}
\newcommand{\eea}{\end{eqnarray}}
\newcommand{\be}{\begin{equation}}
\newcommand{\ee}{\end{equation}}

\newcommand{\bi}{\begin{itemize}}
\newcommand{\ei}{\end{itemize}}

\title{{\Large Dark Matter and LHC:  What is the Connection?}\footnote{\tt{Invited review for Modern Physics Letters A.}}}

\author{   
 Gordon Kane\footnote{\tt{gkane@umich.edu}}, 
 and Scott Watson\footnote{\tt{watsongs@umich.edu}}
\\ Michigan Center for Theoretical Physics\\
University of Michigan, \\ Ann Arbor, Michigan 48109, USA}

\abstract{
We review what can (and cannot) be learned if dark matter is detected in
one or more experiments, emphasizing the importance of combining LHC
data with direct, astrophysical and cosmological probes of dark matter.
We briefly review the conventional picture of a thermally produced WIMP relic density and its connection with theories of electroweak symmetry breaking.
We then discuss both experimental and theoretical reasons why one might generically expect this picture to fail.  If this is the case, we argue that a combined effort bringing together all types of data -- combined with explicitly constructed theoretical models -- will 
be the only way to achieve a complete understanding of the dark matter in our universe and become confident that any candidate actually provides the relic density.}

\preprint{}

\begin{document}

\section{Introduction}
One of the most compelling hints for physics beyond the standard model is the cosmological observation that nearly a quarter of our universe consists of cold dark matter. In the next few years, the Large Hadron Collider (LHC) shows the promise of producing these elusive particles and possibly measuring their microscopic properties \cite{Brhlik:2000dm,list}. This will be challenging and using LHC observations to reconstruct a complete theory of cosmological dark matter could prove even more challenging, if possible at all. New
information will soon come from PAMELA, GLAST, and other experiments
and the interpretation of the reported DAMA annual modulation effect may
be clarified.

One reason is because before we can claim that we know what the dark matter actually is, it will be essential to calculate the relic density of each candidate and compare the total with 
the cosmological relic density determined from the dark matter's gravitational influence -- because it is impossible to measure the relic density directly. 
This calculation will require a detailed knowledge of {\em all} dark matter particles that contribute to the relic density, their interactions, and a knowledge of the cosmological expansion history.
For example, we already know that neutrinos make up roughly a percent of the dark matter, and most complete theories have in addition to massive neutrinos also Weakly Interacting Massive Particles (WIMPs) and axions, so naively there is no convincing reason to expect one form to entirely dominate, as first emphasized in \cite{Brhlik:2000dm}.  
In addition, to calculate the relic density of dark matter candidates we must know how they annihilate as the universe cools, so we can calculate the number left today.  Which diagrams are important depends on the particle interactions.  For example, if the dark matter particles of interest are the lightest superpartner it could behave like the partners of the W boson (wino, $\tilde{W}$), the Higgs boson (higgsino, $\tilde{h}$), the photino $\tilde{\gamma}$, or the sneutrino $\tilde{\nu}$, etc.  These interact differently, and annihilate differently, so it is necessary to know the relevant composition and couplings, which means measuring them, perhaps in the context of a theory that allows some to be calculated.  Each direct, indirect and collider channel can contribute to this information.  
 It is also necessary to know the cosmological history of the universe from the end of inflation until the relic density is no longer changing. Of course, as soon as there are candidates there will be acclaim for the discovery of dark matter, but the question is too important to ignore the need for a proper calculation of the relic density.
 
 In the remainder of the paper we present a review of dark matter, emphasizing the issue of reconstructing a complete picture from both the particle physics and cosmological perspectives -- referring the reader to \cite{Bertone:2004pz} for a more general review, particularly of candidates and assuming an entirely thermal history.
In Section \ref{section1} we review the cosmological evidence for dark matter and its expected distribution inside our galaxy.  In the next section, after a critical overview of the standard calculation of thermal relic density of dark matter -- noting the numerous assumptions that are involved -- we then discuss how this naturally connects the dark matter with new physics near the scale of electroweak symmetry breaking.  In Section 4 we discuss what we call the ``dark matter inverse problem''.  That is, if we were able to reconstruct signatures of dark matter at LHC (the possibility of which we review) is it possible to also reconstruct the dark matter relic density from this information alone and be convinced that we have obtained a complete understanding of dark matter?  The answer is most likely no, and we will give many reasons -- both phenomenological and theoretical-- why this is not only unlikely, but also not expected from complete and self consistent theories. 
Instead, as we discuss in Section 5, by combining LHC data with other direct and indirect detection experiments it may be possible to attain a complete understanding of dark matter.  
Indeed, this is a particularly exciting time, since a number of experiments such as GLAST and PAMELA should be reporting data soon, many other experiments will come online in the near future, and when these astrophysical probes are combined with collider data and cosmological probes a more complete picture of dark matter will emerge.  In the last section we conclude with  a summary of how one should proceed given a detection of a dark matter candidate at LHC, direct, or indirect detection experiment.

\section{Cosmological Evidence for Dark Matter \label{section1}}
The first hint for the existence of dark matter came from observations of the nearby Coma cluster of galaxies by Fritz Zwicky in 1933 \cite{zwicky}.  Zwicky found that by assuming the galaxies comprising the cluster were in equilibrium, their velocity distribution implied a cluster mass far exceeding that inferred from the luminous matter contained within the cluster. Today, through a number of complementary and more sophisticated techniques, cluster studies suggest a relative abundance of dark matter $\Omega_{cdm} = 0.2$ to $0.3$, where $\Omega_{cdm}= \rho_{cdm} / \rho_c$ is the fraction of the critical density in dark matter, $\rho_c=3 H_0^2 m_p^2=8.10 h^2 \times 10^{-47}$ GeV$^{4}$ is the critical density today, with $h$ the Hubble constant in units of $100$ km$\cdot$s$^{-1}$$\cdot$Mpc$^{-1}$ and $m_p=2.43 \times 10^{18}$ GeV is the reduced Planck mass.

A more precise measure of dark matter can be obtained from less direct observations, such as the temperature anisotropies of the cosmic microwave background (CMB), and the evolution and formation of the large scale structure (LSS) of the universe.  This is because the evolution of density inhomogeneities that eventually grow to form LSS is quite sensitive to the properties of the primordial bath of particles from which they evolve.  At the time the CMB photons last scattered, by mass the particles were primarily composed of dark matter.  Combining probes of the CMB, structure formation, and distance probes such as supernovae the amount of dark matter \cite{wmap} is found to be
\be \label{cdm}
\Omega_{cdm} h^2=0.1143 \pm 0.0034,
\ee
with $h=0.70 \pm 1.3$, implying that the total energy budget of the universe is comprised of a little less than a quarter dark matter.

In addition to determining the dark matter abundance, these observations, along with the above mentioned galaxy and cluster observations, also tell us that the dark matter must be `cold' -- meaning non-relativistic, stable (or at least with a lifetime of order of that of the universe today), and `dark' meaning electrically neutral and non-luminous.  The latter, when combined with constraints from Big Bang Nucleosynthesis, suggests the particles are at most weakly interacting.  Combining all of these cosmological observations, we find that what is expected is a WIMP, that is a Weakly Interacting Massive Particle.

Compared to dark matter on larger scales, our local density and distribution of dark matter has proven more challenging to understand.
However, by combining knowledge of galaxies similar to our own, N-body simulations of structure growth, and other observations one can estimate the local density as
\be
\rho^{local}_{cdm}=0.3 \;  \mbox{GeV} \cdot \mbox{cm}^3, 
\ee
where most estimations agree with this result within better than a factor of two\footnote{See e.g. Section 2.4 of  \cite{Jungman:1995df} for a detailed discussion.}.  The distribution of the dark matter throughout the galaxy is also uncertain, but is usually 
described by a radially symmetric distribution 
\be \label{profile}
\rho(r) = \rho_{\odot} \left(\frac{r_{\odot}}{r} \right)^{\gamma} \left(\frac{1+\left(r_{\odot}/r_s \right)^{\alpha}}{1+\left(r/r_s\right)^{\alpha}}\right)^{\left(\beta-\gamma\right)/\alpha},
\ee
where $r_\odot= 8.5$ kpc is our location from the galactic center, $\rho_\odot$ is the local halo density, and $(\alpha, \beta, \gamma, r_s)$ are the parameters modeling the dark matter profile.  
N-body simulations seem to favor cusped profiles 
at the center of the galaxy such as the Navarro-Frenk-White (NFW) \cite{Navarro:1995iw} model given by the parameter choice $(1,3,1,20)$, 
while dynamical observations of galaxies seem to favor cored profiles of 
the isothermal variety \cite{isotherm} with $(2,2,0,3.5)$.  Current dark matter simulations do not yet include the 
effects of baryons. The ratio of dark matter to baryons is about $5.5$, with most of the dark matter residing in the spherical halo, while baryons dominate the gravitational potential in the center of our galaxy.  For a recent analysis of how different profiles can affect signals for dark matter detection (in the case of neutralinos) see  \cite{arxiv:0807.1508,arxiv:0807.1634}.

In regards to the local density and distribution of dark matter, one would hope for significant improvements in years to come.
This is especially true since we will see that the uncertainties in the local relic density will directly translate into uncertainties in particle experiments aimed at the indirect detection of dark matter.

Finally we mention that some researchers are not convinced that dark matter is particles and have instead suggested alternative theories, such as modifications to gravity.  We think the evidence for clumping, and results such as the Bullet Cluster data \cite{Clowe:2006eq} imply convincingly that the dark matter is one or more new kinds of stable or metastable particles, and proceed on this basis in what follows.

\section{Connection to Particle Dark Matter}

\subsection{Dark Matter as a Thermal Relic \label{sectionthermal}}
We begin this section with a brief review\footnote{For a more detailed treatment we refer the reader to the literature  \cite{rocky,weinbergbook}.} of the standard phenomenological approach to the origin and evolution of dark matter as a thermal relic.  

As the universe expands it cools.  
Thus, we expect at some point in the early universe the temperature exceeded the mass of the dark matter particles and they were in thermal and chemical equilibrium.  In equilibrium, the rate at which particles annihilate in a fixed comoving volume $a^3$ is $n_x a^3 \times n_x \langle \sigma_xv \rangle$, where $n_x$ is the number density of dark matter particles of mass $m_x$ and annihilation cross section $\sigma_x$, and $\langle \sigma_xv \rangle$ is the thermally averaged cross-section and relative velocity of the particles.  In equilibrium, particle annihilations should be balanced by particle pair-creation and so that rate is given by $(n_x^{eq})^2 a^3 \langle \sigma_xv \rangle$, so that in equilibrium the number in a comoving volume is constant.  This is expressed by the Boltzmann equation
\be 
\frac{d{\left(n_x a^3\right)}}{dt}= - a^3 \langle \sigma_xv \rangle  \left[  n_x^2 - \left(n_x^{eq}\right)^2 \right] ,
\ee 
where the first term on the right is dilution due to particle annihilations ($XX \rightarrow \gamma \gamma$), and the second term is the reverse process of particle creation from the thermal bath ($\gamma \gamma \rightarrow XX$). At high temperatures when $T \gg m_x$, we have $n_x^{eq} \sim T^3$ and since $T \sim 1/a$ the last two terms cancel and the particle density simply scales with the expansion. Once the particles become non-relativistic ($m_x \ll T$) then $n_x^{eq} \sim e^{-m/T}$ becomes Boltzmann suppressed and particle production becomes negligible, so that the density of particles rapidly drops due to both the expansion and annihilations. Finally, once the cosmic expansion exceeds the annihilation rate per particle $H \gtrsim n_x \langle \sigma_xv \rangle$, the particles `freeze-out' and their number per comoving volume 
is
\be \label{freeze}
\frac{n_x}{s} =\left. \frac{3 H}{s \langle \sigma_xv \rangle} \right\vert_{T=T_f},
\ee
where all parameters appearing in this expression are to be evaluated at the freeze-out temperature $T_f$ and we have introduced the entropy density $s=(2\pi^2/45) g_\ast T^3 \sim 1/a^3$, which gives a more convenient way to define the comoving frame and $g_\ast$ is the number of relativistic degrees of freedom at the time of freeze-out.  The freeze-out temperature can be found from the number density, since one finds that it closely tracks the equilibrium density near freeze-out.  Thus, at freeze-out $n_x \sim n_x^{eq} \sim e^{-m_x/T_f}$, and the mass to temperature ratio at this time is only logarithmically sensitive to changes in the parameters appearing in (\ref{freeze}).  In fact, for thermally produced dark matter associated with weak-scale physics this ratio is typically $m_x/T=25$, with corrections up to at most a factor of two\footnote{Of course, the Boltmann equation can always be solved numerically and one finds good agreement with the analytic argument given above.  See \cite{rocky} for a more thorough discussion.}.

Assuming no significant entropy production, the number of dark matter particles per comoving volume (\ref{freeze}) will be preserved resulting in a density of dark matter
\bea \label{criticaldensity}
\Omega_{cdm}(T) \equiv \frac{\rho_{cdm}(T)}{\rho_c} &=& \frac{m_x n_x(T)}{\rho_c}= \frac{m_x}{\rho_c}  \, \left( \frac{n_x(T_f)}{s(T_f)} \right) s(T)  \\
&=& \frac{m_x}{\rho_c} \left( \frac{3 H}{s \langle \sigma_xv \rangle}  \right)_{T=T_f} s(T). 
\eea
Making the additional assumption that the universe is entirely radiation dominated at freeze-out so that $H\sim T^2$ and using $s \sim T^3$ we find that the critical density in dark matter evaluated today is
\bea \label{cdm2}
\Omega_{cdm}(T_0)&=&\frac{45}{2 \pi \sqrt{10}}  \left( \frac{s_0}{\rho_c m_p}\right)  \left(  \frac{m_x}{g_\ast^{1/2} \langle \sigma_xv \rangle T_f  } \right), \nonumber \\
&=&\frac{1.97 h^{-2} \times 10^{-9} \,  \mbox{GeV}^{-2} }{\langle \sigma_xv \rangle}, 
\eea
where we have used $\rho_c/s_0=3.6 \times 10^{-9}h^2$ GeV, $g_\ast=10.75$, and we note that $1$ GeV$^{-2} =3.89 \times 10^{-28}$ cm$^2 = 3.89 \times 10^8$ picobarns (pb).
If we now compare this result with the cosmological measurement given in (\ref{cdm}), we find that $\langle \sigma_xv \rangle \sim 1$ pb and thus we expect new physics in the form of dark matter at the weak scale, for which this is a typical $\sigma_x$.  As we will discuss further, this result is quite robust. That the weak scale cross section appears in (\ref{cdm2}) along with the consequences of a high temperature big bang, and gives about the observed relic density, is surprisingly rather insensitive to the particle physics and the expansion history of the universe. 

To close this section, we briefly summarize the assumptions that went into the result (\ref{cdm2}):
{
\bi
\item The WIMPs were at some point relativistic and reached chemical equilibrium.
\item At the time of freeze-out the universe was strictly radiation dominated (all other contributions to the energy density were negligible).
\item Following freeze-out there was no significant entropy production.
\item There were no other late-time sources of dark matter particles (e.g. decays from other particles).
\item There is only one species of dark matter particle and any other new particles are unstable or have significantly larger mass.
\ei}

Given the above assumptions:
{ 
\bi
\item The relic density does not depend on the expansion history, only on the temperature at freeze-out.
\item The relic density does not depend on any high scale physics, only on the low-energy cross-section.
\item The answer is very robust to changes in the cross-section and mass of the particles.
\item When combined with cosmological observations -- we expect new physics at the electroweak scale.
\ei
}
\subsection{Electroweak Symmetry Breaking and WIMPs}
There are many reasons - both cosmological and from particle physics - to expect physics beyond the standard model.
Cosmological examples include inflation, dark matter, and the observed baryon asymmetry.
For example, a successful dark matter candidate must be stable, electrically neutral, massive (non-relativistic), and weakly interacting.
This excludes any known standard model particle.  From the perspective of particle physics, the standard model requires a mechanism to give the W and Z bosons mass.
The higgs mechanism provides an explanation, which in-turn implies a fundamental scalar, the higgs boson.  However, we then require an additional symmetry to protect the higgs from radiative corrections and stabilize the electroweak scale.  Without such an additional symmetry, the standard model remains incomplete.

A complete theory of electroweak symmetry breaking typically requires additional particles and symmetries.
Of course at low energies we know that these symmetries are not realized, and therefore they must be spontaneously broken.
After symmetry breaking, in order that the new particles do not decay to standard model particles (which could lead to violations of conserved quantities - such as baryon number) we usual require that there is a residual discrete symmetry.   Thus, all of the new particles will eventually decay down to the lightest particle that is then protected by the residual discrete symmetry.
\subsubsection{Supersymmetry and the LSP}
As an example, consider Supersymmetry (SUSY) as the new symmetry providing a complete theory of electroweak symmetry breaking.
In this case, R-parity or a generalization is the discrete symmetry and this protects the Lightest SUSY Particle (LSP) from decay.  Because this particle is stable, weakly interacting, and has a mass near the electroweak scale, this naturally provides a WIMP candidate.

Depending on the details of SUSY breaking there can be different LSPs.  One of the most typical in models is the neutralino.  In the minimal SUSY extension of the standard model (MSSM) there are two higgs bosons, with neutral components $H_u$ and $H_d$, which give mass to the up-type and down-type fermions, respectively.  Their superpartners the higgsinos $\tilde{H}_u$ and $\tilde{H}_d$, along with the superpartners of the electroweak gauge bosons, the bino $\tilde{B}$, wino $\tilde{W}$ mix to form mass eigenstates called neutralinos following electroweak symmetry breaking.
The lowest mass eigenstate of the neutralino mass matrix,  $\tilde{\chi}$, is then the LSP and therefore a natural candidate for WIMP dark matter.

\section{Dark Matter Inverse Problem}
We have seen that by combining the result for the thermally produced dark matter abundance (\ref{cdm2}) along with cosmological observations (\ref{cdm}), the  
suggestion of new physics near the scale of electroweak symmetry breaking emerges.  It is then quite interesting that an entirely different reason, demanding theoretical consistency of electroweak symmetry breaking, also leads to an expectation of new physics near a TeV.  In fact, it is precisely the latter that generated much interest in the LHC.  
Given that LHC will probe these energies one may hope that, since the relic density (\ref{cdm2}) depended primarily on the thermal averaged cross-section and velocity, from LHC we might determine the relic density of cosmological dark matter.  In particular, because WIMPs are non-relativistic at freeze-out, $\langle \sigma_xv \rangle$ has very weak dependence on the velocity and we could imagine measuring the cross-section at LHC to determine the relic density of the cosmological dark matter.  In this section we want to discuss a number of reasons why this approach is overly optimistic and why relying on LHC alone is unlikely to be enough to obtain a complete understanding of dark matter, even though LHC data will probably be essential for learning what is needed to successfully compute the relic density.

\subsection{Dark Matter at LHC}
The LHC is a proton-proton collider, which will reach a center of mass energy of at least $14$ TeV.  However, because most events of interest will come from the scattering of quarks and other constituents of the protons, which carry at most around $10\%$ of the energy, the probing power of LHC will be around $2$ TeV.  For WIMPs, if we assume that their interactions depend on only weak-scale couplings, rather than strong ones, the cross sections are smaller and the masses probed directly are even less, perhaps up to a few hundred GeV.  Their expected mass is around $100$ GeV, and so it is likely they are produced at LHC, both directly and in decays of heavier parent particles. 
\begin{figure}[ph]
\centerline{\psfig{file=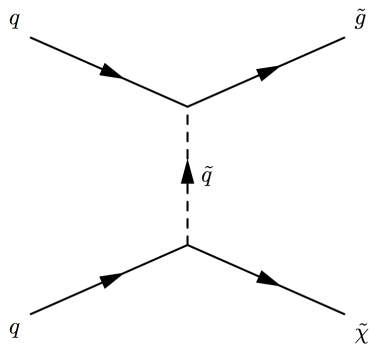,width=2.7in}}
\caption{.\protect\label{fig1} A simple example of a possible process where quarks give rise to a gluino in the final state plus missing energy in the form of an LSP.}
\end{figure}
However, there are obstacles to demonstrating the detection of WIMPs at LHC.  Firstly, the lifetime of particles in the detector will be around $10^{-8}$ s, whereas to confirm cosmological dark matter we need to measure the lifetime to exceed the current age of the universe $10^{17}$ s.
This can only be overcome by also observing the candidate directly or indirectly in astrophysical experiments, comparing the candidates from the the two observations, and then finding agreement on their mass and couplings.  
Another difficulty is that because the WIMPs are weakly interacting they will pass through the detector and their presence will only be inferred from missing transverse momentum in the events.  
Therefore, to understand the properties of WIMPs we will need to effectively reconstruct the events and study the other associated decay products.
\begin{figure}[ph]
\centerline{\psfig{file=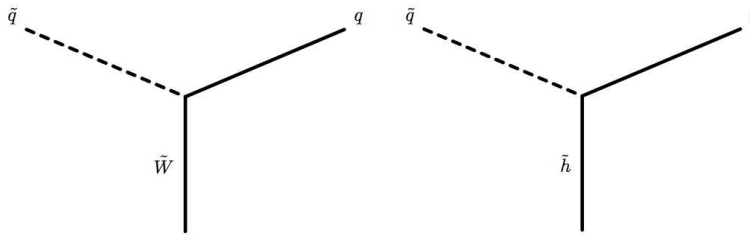,width=4.0in}}
\caption{.\protect\label{fig2} In the case of the neutralino, vertices like those shown above can help distinguish the nature of the neutralino, when they are part of production or decay processes.  That is, if the neutralino is primarily wino-like then squark-quark vertices like the one above (left) will be present, whereas the same vertex for the higgsino above (right) is strongly suppressed since the coupling is proportional to the small fermion masses.}
\end{figure}
We can take SUSY as an example, where quark and gluon scattering will lead to production of squarks, gluinos, neutralinos, and charginos.
These superpartners will eventually decay down to energetic Standard Model particles, such as quarks and leptons, and the remaining missing energy will be in the form of LSPs (WIMPs) that escape the detector.   A simple example of a possible process is shown in Figure \ref{fig1}, where quarks give rise to a gluino in the final state plus missing energy in the form of an LSP.  Diagrams such as these have backgrounds and small rates, so they are not discovery channels.  We will not measure the neutralino directly, but instead can hope to deduce its properties after gluino and squarks masses and branching ratios are known.  In particular, it will be important to perform a careful analysis of the data looking for signatures that can help distinguish the LSP from other candidates.  In the case of the neutralino, vertices like those shown in Figure \ref{fig2} can help distinguish the nature of the neutralino, when they are part of production or decay processes.  That is, if the neutralino is primarily wino-like then squark-quark vertices like the one in Figure \ref{fig2} (left) will be present, whereas the same vertex for the higgsino (Figure \ref{fig2} right) is strongly suppressed since the coupling is proportional to the small fermion masses.  Noting this, if we did indeed see an event like that of Figure \ref{fig1}, we could be sure that the neutralino is not primarily higgsino.   As another example, if the neutralino is primarily wino-like then the chargino (the mass eigenstate of the charged superpartners of the $SU(2)$ gauge bosons) will be nearly degenerate in mass with the neutralino. This implies that tri-lepton events will be highly suppressed leading to another indication of a wino-like neutralino.  

Combining clues like these with predictive models of electroweak symmetry breaking should not only help in understanding the new physics, but also make an understanding of particle dark matter possible.  In fact, in special cases with models that are highly constrained strong bounds have already been placed on particle dark matter  \cite{Arnowitt:2008bz,Nojiri:2005ph}, and LHC data can go far toward determining the relic density if the models are correct. 

However, in general deducing the properties of the LSP from data could prove quite formidable, especially given the large parameter space of theories beyond the standard model. That is, given a signal at LHC there can be many regions of parameter space (or even many theories) that predict the signature -- this is the LHC inverse problem \cite{ArkaniHamed:2005px}.
One way to address this is to combine LHC data with additional astrophysical probes (discussed in Section \ref{indirect}), such as direct detection and indirect detection through observations of annihilation products to break degeneracies (see e.g. \cite{Bourjaily:2005ax,Altunkaynak:2008ry}).   For example, it was shown in \cite{ArkaniHamed:2005px} that within the MSSM there can be many regions of parameter space that give rise to scalar, LSP, and NLSP masses that are indistinguishable at LHC.  However, it was pointed out in \cite{Bourjaily:2005ax,Altunkaynak:2008ry} that these models give rise to vastly different relic densities and direct and indirect detection signatures for the LSP.  
Another source of uncertainties arises from the complex phases appearing in the softly broken SUSY Lagrangian.  Even if these are taken to vanish, it was first stressed in 
\cite{Brhlik:2000dm} that this can still introduce a large uncertainty of around six into the critical density of dark matter in LSPs. 
 
By reconstructing events at colliders such as LHC we might obtain the LSP mass within sufficient accuracy. 
This will still have to be compared with direct detection \cite{Green:2008rd} and astrophysical probes of dark matter to ensure that it is indeed WIMPs making up the cosmological dark matter.  
Ultimately, because the LHC signatures will depend on the composition and properties of the dark matter -- though not directly -- we are hopeful that systematic study will allow deducing the LSP properties needed to calculate the relic density. In \cite{Baltz:2006fm} it was suggested that  a linear collider (e.g. ILC) is needed. 
Moreover, as we will discuss in the following sections, many of the assumptions leading to the thermal relic abundance (\ref{cdm2}) seem to be at odds with our current theoretical understanding of the early universe, and of actual theories that predict dark matter.

\subsection{Challenges for Reconstructing the Cosmological Relic Density}
In addition to the challenge of reconstructing the properties of dark matter from signatures at colliders, direct, and indirect detection, there are also a number of challenges associated with the relic density of dark matter itself.  These include the fact that the relic density could be comprised of more than one kind of particle, the way in which dark matter is produced, and the expansion history of the universe.

\subsubsection{Other Dark Matter}
One key assumption underlying the connection between the thermal relic abundance (\ref{cdm2}) and LHC, is that the WIMP is a unique dark matter candidate and that its mass is far below the next to lightest particle associated with new physics.  
As an example of the latter, in supersymmetric theories it is common that the next to lightest SUSY particle (NLSP) can be nearly degenerate in mass with the LSP.
If this is the case, not only could the NLSP be mistaken as a stable WIMP (LSP) in the LHC detector -- as the lifetime of a particle in the detector is only $10^{-8}$ s, or the NLSP and its decay products might both be neutral -- but cosmologically, coannihilations \cite{coann} between the NLSP and LSP will significantly reduce the thermal relic density estimated in (\ref{cdm2}).

Another important possibility is that there is more than one type of dark matter.  
Thus, the total dark matter abundance should always be thought of as
\be
\Omega_{cdm}^{total} = \sum_i \Omega_{cdm}^{(i)},
\ee
where the sum is over all contributions to the dark matter energy budget.
In fact, because we now know that neutrinos have mass, we also know that they must make up some part of the dark matter.  
However, we also know that neutrinos are relativistic at the onset of structure formation, i.e. they are  `warm' dark matter, requiring that they must represent a small fraction of the total dark matter.  In fact, combining the recent WMAP5 data with other cosmological observations a bound of  $\Omega_\nu h^2 \lesssim 0.006$ was obtained in \cite{wmap}. Of course, in addition to neutrinos there are a number of other possible contributions to the cosmological dark matter, including axions.  The QCD axion provides an elegant solution to the strong CP problem, and although tightly constrained still remains a viable dark matter candidate for some regions of its parameter space (see e.g. \cite{arxiv:0807.1726}).
It is also expected that additional axions will generically arise at low energies from effective theories with ultraviolet completions in string theory (see \cite{Svrcek:2006yi} and references within). 
Indeed, it will be difficult if not impossible to calculate an axion relic density \cite{Brhlik:2000dm}.  Perhaps the only way we could become convinced axions constituted the dark matter would be if the calculated WIMP relic density were small compared to the experimentally determined value.

\subsubsection{Modified Expansion History at Freeze-out}
For the calculation of the thermal relic density in Section \ref{sectionthermal} one assumption was that the universe was radiation dominated at the time of freeze-out, so that $H \sim T^2$
allowing for the simplification in going from (\ref{criticaldensity}) to  (\ref{cdm2}).  This assumption agrees with the observational predictions of Big Bang Nucleosynthesis  (BBN) occurring at around one second after the Big Bang.  However, there is no cosmological evidence for this assumption prior to the time of BBN.  

There are both theoretical and observational indications that this assumption may be too naive.
Indeed, given the rich particle phenomenology that occurs at energies above the scale of BBN (MeV), we might expect this to complicate the simple picture of a purely radiation dominated universe.
Moreover, relics from early universe phase transitions, such as scalar condensates or rolling inflatons that didn't completely decay, would also be expected to alter the expansion history.

In fact, theories beyond the standard model generically predict the existence of scalar fields. Many of these fields have little or no potential -- so called moduli, so they are often light.  Examples include the sizes and shapes of extra dimensions, or flat directions in the complicated SUSY field space of the scalar partners to standard model fermions.  
In the early universe these moduli will generically be displaced from their low energy minima during phase transitions, such as inflation \cite{Dine:1995kz}.  Energy can then become stored in the form of coherent oscillations forming a scalar condensate.   The cosmological scaling of the condensate depends on which term in the potential is dominant.  For a potential with a dominant term $V \sim \phi^\gamma$ one finds that the pressure depends on the energy density as
\be \label{pressure}
p=\left( \frac{2 \gamma}{2+\gamma} - 1 \right) \rho,
\ee
where $\rho$ scales as
\be
\rho = \rho_0 a^{-6 \gamma / (2+\gamma)}.
\ee

Two examples are a massive scalar with negligible interactions for which $\gamma=2$ and the condensate scales as pressure-less matter $p=0$, whereas if physics at the high scale is dominant -- in the form of non-renormalizable operators -- then $\gamma > 4$ and the condensate evolves as a stiff fluid $p \approx \rho$ for large $\gamma$.
Whatever the behavior of the condensate, if it contributes appreciably to the total energy density prior to freeze-out the abundance (\ref{cdm2}) will be altered.  This is because the presence of addition matter will increase the cosmic expansion rate allowing less time for particle annihilations prior to freeze-out\footnote{Here we have assumed that radiation contributes substantially to the total energy density or that whatever the primary source of energy density it scales at least as fast as radiation.  However, if instead the universe were completely dominated by a massive, non-interacting scalar condensate then this would actually decrease the amount of dark matter.  In either situation, the point is that the standard thermal relic density (\ref{cdm2}) will not give the correct result.}.
The expansion rate at the time of freeze-out is then given by
\be
H_f = H_{rdu} \left(1 + \frac{\rho_\phi}{\rho_r}  \right)^{1/2},
\ee
where $H_{rdu}$ is the expansion rate in a radiation dominated universe and $\rho_\phi$ and $\rho_r$ are the energy density of the scalar condensate and radiation, respectively.
Using that at freeze-out $\rho_r = (\pi^2/30) g(T_f) T_f^4$, $\rho_\phi= \rho_{osc} \left(  T_f /T_{osc} \right)^p$ where $p \equiv 6 \gamma / (2+ \gamma)$, and $\rho_{osc}$ is the energy initially in the condensate which began coherent oscillations at temperature $T_{osc}$ we find that the new dark matter abundance is 
\be
\Omega_{cdm} \rightarrow \Omega_{cdm}  \sqrt{1+r_0 T_f^{2(\gamma-4)/(2+\gamma)}} ,
\ee
where we have used $a \sim 1/T$ for an adiabatic expansion, and the constant 
$$
r_0 \equiv \frac{30}{\pi^2} \left(\frac{\rho_\varphi(T_{osc}) }{ g(T_f) T_{osc}^{6\gamma/(2+\gamma)} }\right),
$$ 
where $g(T_f)$ is the number of relativistic degrees of freedom at freeze-out.
In practice, typically one finds that for moduli in the early universe $r_0 \gg 1$ \cite{Dine:1995kz,Acharya:2008bk}.
We see that especially for high energy effects in the potential this can have a significant effect on the resulting relic density.
As a simple example, if we consider a massive scalar with negligible interactions ($\gamma=2$) displaced after a period of inflation we expect $\rho_\varphi(T_{osc}) \simeq m_\varphi^2 m_p^2$ so that $r_0 \simeq (m_\varphi m_p)^2 / (g(T_f) T_{osc}^3) \gg 1$ leading to a large enhancement of the relic density.
One can also show that there is a significant effect for scalars which are dominated by their kinetic terms (e.g. kination models \cite{Salati:2002md,Catena:2004ba,Chung:2007vz,Chung:2007cn} ), which 
behave like the stiff fluid models discussed above (i.e. $p=\rho$).   In fact, this modification to the expansion history was considered in \cite{Grin:2007yg}, where it was shown that this would loosen constraints on axionic dark matter.
In these examples, the relic density is found to be enhanced compared to that of a purely radiation dominated universe.
Of course scalar condensates are not the only additional sources of energy one might expect in the early universe and it is important to note that any additional, significant component will alter the standard thermal abundance of the cosmological dark matter in a way similar to that discussed for scalars above.   
\subsubsection{Late Production of Dark Matter and Entropy}
Two more crucial assumptions that went into the dark matter abundance (\ref{cdm2}) were that there were no other sources of dark matter and/or entropy production following freeze-out. An example of how this can fail is if there is a late period of thermal inflation \cite{Lyth:1995ka}, which has been argued to be quite natural and necessary for resolving issues with some models coming from string compactifications (see e.g. \cite{Conlon:2007gk}). Another example is provided by the condensate formation we discussed above. 
That is, because the moduli have very weak couplings -- typically of gravitational strength -- the condensate will decay late producing additional particles and entropy.  This decay must occur before BBN, which requires the modulus to have a mass larger than around $10$ TeV in order to avoid the so-called cosmological moduli problem
 \cite{Coughlan:1983ci,Ellis:1986zt,deCarlos:1993jw,Banks:1993en}.  If the condensate contributes appreciably to the total energy density at the time it decays it will not only produce relativistic particles -- and significant entropy -- but could also give rise to additional dark matter particles.  The former will act to reduce the thermal relic density of dark matter particles $\Omega_{cdm} \rightarrow \Omega_{cdm} \left(  T_r / T_f \right)^3$, where $T_r$ is the temperature after the decay and $T_f$ is the freeze-out temperature of the dark matter particles.  As an example, for a $10$ TeV scalar the decay to relativistic particles will `reheat' the universe to a temperature of around an MeV, whereas a $100$ GeV WIMP freezes out at a temperature near a GeV.  Thus, the scalar decay will dilute the preexisting relic density in dark matter by a factor of about $(T_r/T_f)^3 \simeq 10^9$.  

As we have mentioned, in addition to the scalar decaying to relativistic particles it could also decay to WIMPs below their freeze-out temperature.  In this case there are two possible results for the relic density, depending on the resulting density of the WIMPs that are produced \cite{Moroi:1999zb}.   If the number density of WIMPs exceeds the fixed point value
\be
\frac{n_x}{s} =\left. \frac{3 H}{s \langle \sigma_xv \rangle} \right\vert_{T=T_r},
\ee
then the WIMPs will quickly annihilate down to this value, which acts as an attractor.  It is important to note that the fixed point value is evaluated at the time of reheating, in contrast to the freeze-out result (\ref{freeze}).  The other possibility is that the WIMPs produced in the decay do not exceed the fixed point value.  In this case their density is just given by $n_x \sim B_x n_\varphi $, where $B_x$ is the branching ratio for scalar decay to WIMPs and $n_\varphi $ is the number density of the scalar condensate.
We see that in both these cases the thermal relic density (\ref{cdm2}) would give the wrong answer for the true abundance of dark matter, unless the entropy diluted thermal density of dark matter still manages to exceed the amount coming from the scalar decay.  We see that in the case of fixed point production the standard thermal relic density
is enhanced by
\be
\Omega_{cdm} \rightarrow \Omega_{cdm} \left( \frac{T_f}{T_r} \right),
\ee
which for the example of a $10$ TeV scalar results in an overall enhancement of a factor of $1000$.  Fixing the relic density by the cosmological data (\ref{cdm}) implies that  particles need a larger cross section in order to get the right amount of dark matter.  For example, in the case of neutralino dark matter, Winos and Higgsinos  annihilate well and have been seen as giving too little dark matter given a thermal history.  However, in the theoretically constructed models of \cite{Moroi:1999zb,Acharya:2008bk} it is found that Winos, HIggsinos, or some mixture can yield the right amount of dark matter due to non-thermal production, which results naturally by requiring consistency of the theory.

It is important to note that even though the naive thermal freeze-out calculations no longer determines the relic density, the answer is still given in terms of the weak scale cross section, and gives a result of the correct order of magnitude for WIMPs with masses of order $100$ GeV. The dark matter scale and the electroweak symmetry breaking scale still remain related, and the ``WIMP Miracle'' remains valid.

\subsubsection{Expectations from a Complete Theory (Top-Down Approach) \label{topdown}}
In the previous sections we have discussed a number of ways that the standard thermal approach to dark matter could fail.
We have seen that the bottom-up approach of thermally produced 
dark matter is both simplistic and independent of the details of UV physics and the cosmic expansion history of the early universe.
This focuses us on the major question, {\em How generic are the effects undermining the thermal relic density?}   

In order to answer this question, we need model independent notions of what to expect in the early universe and from underlying theories.
This is of course a challenging task and part of the theoretical appeal of the thermally produced dark matter paradigm.
A possible framework in which to address these questions is string / M theories.
Knowledge of string based model building has drastically improved over the past decade.
Theorists are now able to construct quasi-realistic models, which contain some generic features that are expected to appear in any successful model. 

Key to such constructions is the stabilization of light scalars -- or moduli -- appearing in the low-energy theory.
These moduli parameterize the compactification of any extra dimensions and describe the location and orientation of any strings and/or branes that are present.
Stabilization of these scalars is then achieved by various non-perturbative effects such as the condensation of gauginos that become strongly coupled in a hidden sector \cite{Nilles:1982ik}, and/or turning on fluxes \cite{Giddings:2001yu} -- generalized Maxwell fields. This stabilization is supplemented with either additional branes \cite{Kachru:2003aw} or additional hidden sector matter fields \cite{Lebedev:2006qq} in order to obtain a nearly Minkowski or de Sitter vacuum today.  
All of these properties are generic in classes of string theories.  One would have to go to considerable trouble to construct theories without them, if that was even possible at all.

Explicit string constructions that accomplish the stabilization of moduli, and provide a realistic vacuum can be quite intricate.  However, demanding that such constructions respect the electroweak hierarchy and give a dS vacuum with softly broken SUSY leads to many similarities between models.  In particular, it was recently shown in \cite{Acharya:2008bk} that such constructions can make concrete predictions about the nature of dark matter.  In fact, in many string based models that have gravity or anomaly mediated SUSY breaking and address the electroweak hierarchy, one can show that there will exist at least one light scalar with mass $m_\varphi \simeq m_{3/2}$, where $m_{3/2} \simeq 10-100$ TeV is the gravitino mass \cite{ps}.  Given our discussion above, this implies that dark matter in such scenarios should result primarily from late-time production.  

In addition to the string constructions above, late-time production has already been seen to result from a number of other works including:  anomaly mediated SUSY breaking models \cite{Moroi:1999zb}, String and M-theory constructions \cite{Acharya:2008bk}, models with low reheat temperature \cite{Giudice:2000ex}, from decay of cosmic strings \cite{Jeannerot:1999yn,Cui:2007js}, and from relics of Affleck-Dine baryogenesis (see \cite{Dine:2003ax} and references within).  In all of these cases the above picture of late-time production is the relevant one, and it is difficult to see how thermally produced dark matter can play a dominant role.  

\section{Other Probes of Dark Matter \label{indirect}}
We have seen that collider data alone is very unlikely to provide us with a complete picture of cosmological dark matter.  In particular, we have seen that there  many theories yielding the same experimental signatures at LHC and reconstructing the relic density from LHC observations alone will not be unique\footnote{For a recent study of using the observed relic density and assuming a thermal history to remove LHC degeneracies see \cite{Altunkaynak:2008ry}.}. This makes cosmological and astrophysical probes of dark matter an essential complementary approach, capable of breaking these degeneracies and helping us develop a more complete picture of cosmological dark matter.  In this section we will briefly discuss direct and indirect detection of dark matter referring the reader to the recent review \cite{Hooper:2008sn} for more details.

\subsection{Direct Detection}
Direct detection experiments measure the properties of WIMPs as they elastically scatter with atomic nuclei.  
These experiments use the rate of scattering, along with the estimated local mass density in the halo ($\rho_{halo} \simeq 0.3$ GeV/cm$^{3}$) and the WIMP velocity\footnote{The WIMPs are assumed to be at rest in the galactic halo and the velocity is the result of the motion of the solar system through the halo.} ($v_w \simeq 230\pm 20$ km/s) to constrain the interaction cross-section as a function of WIMP mass.  Direct detection has the advantage that measurements are made locally.  This is in contrast to indirect detection methods where uncertainties can arise from our lack of knowledge of the distribution of dark matter throughout part or all of the halo and from those associated with particle propagation throughout the galactic magnetic field.  Instead, barring any local inhomogeneities (which seem quite unnatural \cite{astro-ph/0201289}), the primary source of uncertainties for direct detection arise from the WIMP-nucleon and WIMP-nucleus cross sections (introducing a factor of roughly four or more in some situations \cite{Hooper:2008sn}).   In general, WIMP-nucleon scattering will receive contributions from both spin dependent and spin independent interactions.  For spin-dependent interactions the cross section scales as $J(J+1)$, whereas for spin-independent scattering the cross section scales as the square of the atomic mass, making it advantageous to use heavy nuclei.  The strongest current limits on the latter come from XENON \cite{arxiv:0706.0039} and CDMS \cite{astro-ph/0509259}, which bound a $100$ GeV mass WIMP to have a WIMP-nucleon interaction cross section of less than $10^{-7}$ pb.  Given a knowledge of the dark matter velocity within $10 \%$ error, it is projected that a Xenon or similar target should be capable of measuring the WIMP mass to within $20 \%$ \cite{Hooper:2008sn}.   Comparing this mass to one(s) deduced from collider and other indirect detection methods will then give an important check on which dark matter candidate is being observed and also help reduce errors in their velocity distribution.   We also note that the situation with the
recently reported DAMA collaboration annual modulation is not yet clear.

\subsection{Indirect Detection}
In addition to direct detection methods it is also possible to deduce WIMP properties indirectly by observing anti-matter, synchrotron radiation, neutrinos, and gamma rays coming from WIMP annihilations.  
This is an enticing possibility, especially given past balloon born experiments such as HEAT (positrons) \cite{HEAT}, EGRET (Gamma rays) \cite{EGRET}, and the space shuttle experiment AMS-01 \cite{ams}, which some have speculated indicate fluxes in excess of astrophysical backgrounds at energies that are relevant  for WIMP annihilations.
We have seen that WIMPs cease to annihilate in the early universe, but inside galaxies densities can reach levels where dark matter will again annihilate. 
For particles produced from such events, the number of particles produced per unit time, energy and volume is given by 
\be \label{source}
Q(E,\vec{x}) = \frac{1}{2} \langle \sigma_xv \rangle \left(  \frac{ \rho_x(\vec{x}) }{m_x} \right)^2 \sum_i B_i \, \frac{dN^i}{dE},
\ee
where $E$ is the produced particles kinetic energy, $\langle \sigma_xv \rangle$ is the annihilation cross section at rest for the non-relativistic dark matter, $B_i$ and $dN^i/dE$ are the branching ratios and fragmentation functions, and the sum is taken over all annihilation channels.  We see that the amount of particles produced depends quadratically on the dark matter number density $n_x(\vec{x}) = \rho_x / m_x$ at the location of the annihilation.  The density of WIMPS is typically modeled by a spherically symmetric profile (\ref{profile}) as discussed in Section \ref{section1}, and we see that for various choices of halo profile and clumpiness we can get different predictions for the amount of produced particles.  In addition to this uncertainty, once the particles are produced they will undergo a random walk through the galactic magnetic field until some of them will finally pass through balloon or satellite detectors. 

\subsubsection{Anti-Matter}
 For positrons the diffusive process will result in a rapid reduction of their energy due to effects such as scattering off of cosmic microwave background photons and synchrotron radiation.  In fact, the latter can be used to place a constraint on the self annihilation cross section of the dark matter particles \cite{Hooper:2007kb,arxiv:0807.1508}.  Because the positrons will quickly loose energy this means that most detected flux will come from nearby annihilations (within a few kpc).
However, anti-protrons can travel from much farther away and when produced outside the galactic plane can then pass through the interstellar medium creating additional secondary signals.  Although both of these effects can lead to additional uncertainties one can make use of other cosmic ray spectra, such as Boron and Carbon, to get better constraints on the diffusion process and backgrounds \cite{astro-ph/0101231}.  
Recent work \cite{arxiv:0807.1508} has concluded that constraints allow a window for wino or higgsino LSPs with mass of order 200 GeV, as suggested by the HEAT \cite{HEAT} and AMS-01 \cite{ams}
data, and could be confirmed soon by PAMELA and GLAST.

\subsubsection{Gamma Rays}
Gamma rays in the energy range of those produced by annihilating dark matter propagate directly to us without undergoing diffusive processes and therefore do not suffer from the same uncertainties as anti-matter.  This also means that detected gamma ray fluxes can give important information about the location of the annihilating dark matter (e.g. angle relative to the galactic plane).
In particular, if PAMELA reports a signal in the positron spectrum it would imply wino or
higgsino (or a mixture of them) dark matter, and GLAST could confirm
such a signal soon after \cite{arxiv:0807.1508}.
One challenge for this method of detection is the large flux of gamma rays coming from the galactic center, which seems to be indicative of astrophysical processes rather than dark matter annihilations \cite{Hooper:2008sn}.  However, given the large density of dark matter expected at the galactic center it is not only a challenge to get a better understanding of the astrophysical background there, but also of the distribution and cuspiness of dark matter.  
Current and upcoming experiments, such as GLAST \cite{GLAST}, will not only help probe the astrophysical background in gamma rays, but will also be capable (along with ground based telescopes) of detecting gamma rays coming from WIMP annihilations.

\subsubsection{Neutrino Telescopes}
Another possible for indirect detection of WIMPs results if they are captured by the Sun and then annihilate with nuclei resulting neutrinos that can then escape to Earth.
In order to detect these neutrinos over the background an event rate of 10-100 events per square-kilometer per year is required.
Experiments such as Amanda, Ice-Cube, and Antares are expected to significantly improve the current upper bound, which is set by Super-Kamionkande \cite{hep-ex/0404025}.
However, for kilometer-scale neutrino telescopes for signal to be detectable the spin-independent cross section for the WIMP-nucleon scattering must be at least $10^{-6}$ pb, which is already ruled out by Xenon \cite{arxiv:0706.0039} and CDMS \cite{astro-ph/0509259}. Therefore, one must rely on WIMP annihilations that scatter with nuclei in the sun via spin-dependent interactions.  For WIMPS such as neutralinos this could be an important source of detection, since neutralinos with a significant higgsino component will couple strongly through spin-dependent interactions, whereas it could be undetectable in other experiments depending on parametric choices in the MSSM.

\section{Conclusions and Outlook}
Both cosmological observations and theoretical consistencies of electroweak symmetry breaking seem to suggest the presence of WIMPs.
The expectation of data coming from LHC makes this an exciting time to ask whether these two notions of dark matter are one and the same -- or if the story is perhaps more complicated.
We have seen in this review that theory would suggest the latter, but that by combining data from colliders with that coming from cosmological and astrophysical probes of dark matter a complete understanding of dark matter can probably be achieved.  In fact, one is tempted to compare this situation with the concordance approach of modern precision cosmology where one experiment alone (such as WMAP) is not enough to overcome uncertainties, but when combined with other probes (such as supernovae, baryon acoustic oscillations, galaxy surveys, etc.) one can obtain a more constrained and precise fit to the cosmological parameters.  Thus, by combining the data from different experiments in this way -- along with explicit and complete theoretical models that not only predict specific dark matter candidates, but also the cosmological expansion history -- we can hope to obtain a complete understanding of dark matter.  
\\

\noindent {\bf How should one proceed once a dark matter candidate is established?} \\
One good path is to calculate the relic density in models that are sufficiently complete to have a cosmological history, and accommodate electroweak symmetry breaking.  Presumably several different models would emerge as consistent
with dark matter and LHC data, but all of them would have other predictions and tests that would allow one class of them to eventually become convincing.

Another approach would be to calculate the relic density as if the history were thermal (using the experimentally inferred self annihilation cross section).  Then there are several possibilities (as discussed above):
\bi
\item The relic density could come out about right (i.e., agree with the cosmological abundance determined by e.g. WMAP), in which case it would belong to the previous paragraph.  
\item The relic density could come out smaller than the WMAP number, in which case either:
\bi
\item There is another source of dark matter, e.g. axions or some heavy hidden sector matter.
\item And/Or, the dark matter was non-thermally produced yielding an increased relic density.
\item And/Or, at the time the dark matter was produced the cosmic expansion was not entirely radiation dominated, e.g. kination.
\ei 
\item The relic density could come out larger than the WMAP number, in which case either:
\bi
\item  There was non-thermal production yielding less than the thermal relic density.
\item  And\footnote{Note:  As we discussed above, at the time non-thermal production occurs (e.g. from decay of a heavy scalar) this will produce substantial entropy.  This will dilute the thermal abundance, which will also contribute to the total relic density.  One could imagine more complicated scenarios where all of these factors play a role and the relic density of dark matter gets both thermal and non-thermal contributions (as is expected for example in the case of axions \cite{arxiv:0807.1726}).}/Or, there were additional sources of entropy after thermal freeze-out (e.g. decays of heavier particles, such as scalars (moduli) or gravitinos). 
\item And/Or the cosmic expansion was modified after thermal freeze-out of the dark matter (e.g. thermal inflation).
\ei
\ei

In all cases, we will be learning, amazingly, about the cosmological history of the universe, from LHC, direct, and indirect detection of dark matter.

\section*{Acknowledgments}

We would like to thank Bobby Acharya, Konstantin Bobkov, Dan Chung, Sera Cremonini, Phill Grajek, Mark Hertzberg, Nemanja Kaloper, Piyush Kumar, Siew-Phang Ng, Alexey Petrov, Dan Phalen, Aaron Pierce, Jing Shao, and Lian-Tao Wang for useful discussions.
The research of G.L.K. and S.W. is supported in part by the US Department of Energy.  S.W. would also like to thank the Stony Brook physics department
and the 2008 Simons' Workshop in Mathematics and Physics for hospitality and financial assistance.

\end{document}